\documentclass[11pt]{article}
\usepackage{amssymb,amsmath,amsfonts}
\usepackage{graphicx}
\usepackage{graphics}
\usepackage{eepic,epsfig}

\begin{document}

\title{Vacuum polarization induced by a cosmic string \\
in anti-de Sitter spacetime}
\author{E. R. Bezerra de Mello$^{1}$\thanks{%
E-mail: emello@fisica.ufpb.br}\, and A. A. Saharian$^{1,2}$\thanks{%
E-mail: saharian@ysu.am} \\
%EndAName
\\
\textit{$^1$Departamento de F\'{\i}sica-CCEN, Universidade Federal da Para%
\'{\i}ba}\\
\textit{58.059-970, Caixa Postal 5.008, Jo\~{a}o Pessoa, PB, Brazil}\vspace{%
0.3cm}\\
\textit{$^2$Department of Physics, Yerevan State University,}\\
\textit{1 Alex Manoogian Street, 0025 Yerevan, Armenia}}
\date{ }
\maketitle

\begin{abstract}
In this paper we investigate the vacuum expectation values (VEVs) of the
field squared and the energy-momentum tensor associated with a massive
scalar quantum field induced by a generalized cosmic string in $D$%
-dimensional anti-de Sitter (AdS) spacetime. In order to develop this
analysis we evaluate the corresponding Wightman function. As we shall
observe, this function is expressed as the sum of two terms: the first one
corresponds to the Wightman function in pure AdS bulk and the second one is
induced by the presence of the string. The second contribution is finite at
coincidence limit and is used to provide closed expressions for the parts in
the VEVs of the field squared and the energy-momentum tensor induced by the
presence of the string. Because the analysis of vacuum polarizations effects
in pure AdS spacetime have been developed in the literature, here we are
mainly interested in the investigation of string-induced effects. We show
that the curvature of the background spacetime has an essential influence on
the VEVs at distances larger than the curvature radius. In particular, at
large distances the decay of the string-induced VEVs is power-law for both
massless and massive fields. The string-induced parts vanish on the AdS
boundary and they dominate the pure AdS part for points near the AdS horizon.
\end{abstract}

\bigskip


\bigskip

\section{Introduction}

Symmetry breaking phase transitions in the early universe have several
cosmological consequences and provide an important link between particle
physics and cosmology. In particular, different types of topological objects
may have been formed by the vacuum phase transitions after Planck time \cite%
{Kibble,V-S}. Among them the cosmic strings are of special interest.
Although recent observational data on the cosmic microwave background have
ruled out cosmic strings as the primary source for primordial density
perturbation, they are still candidate for the generation of a number of
interesting physical effects such as gamma ray bursts \cite{Berezinski},
gravitational waves \cite{Damour} and high energy cosmic rays \cite%
{Bhattacharjee}. Recently, cosmic strings have attracted renewed interest
partly because a variant of their formation mechanism is proposed in the
framework of brane inflation \cite{Sarangi}-\cite{Dvali}.

The gravitational field produced by a cosmic string may be approximated by a
planar angle deficit in the two-dimensional sub-space. In quantum field
theory the corresponding non-trivial topology induces non-zero vacuum
expectation values for physical observables. The analysis of the vacuum
polarization effects associated to scalar and fermionic fields have been
done in \cite{scalar}-\cite{Site11} and \cite{ferm}-\cite{Beze08f},
respectively, for the geometry of an idealized cosmic string on background
of flat spacetime. For curved backgrounds additional effects appear due to
the bulk gravitational field. Recently the combined effects of non-trivial
topology and background curvature on the local characteristics of the scalar
and fermionic vacua have been investigated in \cite{StringdS,StringdS1} for
a cosmic string in de Sitter spacetime. It has been shown that, depending on
the curvature radius of de Sitter spacetime, two regimes are realized with
monotonic and oscillatory behavior of the vacuum expectation values at
distances larger than the de Sitter curvature radius. In this paper we want
to continue along similar line of investigation analyzing the vacuum
polarization effects associated with a massive scalar quantum field in a
high dimensional anti-de Sitter (AdS) spacetime in the presence of a cosmic
string. Recently, the geometry of a cosmic string in the background AdS
spacetime has been discussed in \cite{Ghe1,Cristine}. In \cite{Cristine},
AdS/conformal field theory (CFT) correspondence was used for the calculation
of the Green function for a scalar field living on the cone.

AdS spacetime is remarkable from different points of view. The early
interest in this spacetime was motivated by the question of principal nature
related to the quantization of fields propagating on curved backgrounds. The
presence of both regular and irregular modes and the possibility of
interesting causal structure lead to a number of new phenomena. The
importance of this theoretical work increased when it was discovered that
AdS spacetime generically arises as a ground state in extended supergravity
and in string theories. Further interest in this subject was generated by
the appearance of two models where AdS geometry plays a special role. The
first model, the AdS/CFT correspondence (for a review see \cite{Ahar00}),
represents a realization of the holographic principle and relates string
theories or supergravity in the AdS bulk with a conformal field theory
living on its boundary. It has many interesting consequences and provides a
powerful tool to investigate gauge field theories. The second model is a
realization of a braneworld scenario with large extra dimensions and
provides a solution to the hierarchy problem between the gravitational and
electroweak mass scales (for reviews on braneworld gravity and cosmology see
\cite{Brax03,Maar10}). In this model the main idea to resolve the large
hierarchy is that the small coupling of 4-dimensional gravity is generated
by the large physical volume of extra dimensions. Braneworlds naturally
appear in the string/M theory context and provide a novel setting for
discussing phenomenological and cosmological issues related to extra
dimensions.

Among the most important characteristics of the vacuum state are the vacuum
expectation values (VEVs) of the field squared and the energy-momentum
tensor. In the present paper we evaluate these VEVs for a massive scalar
field with general curvature coupling parameter in the geometry of a cosmic
string on background of AdS spacetime. Though the corresponding operators
are local, due to the global nature of the vacuum state these quantities
carry an important information about the topology of the spacetime. In
addition to describing the physical structure of the quantum field at a
given point, the energy-momentum tensor acts as the source in the Einstein
equations and therefore plays an important role in modelling self-consistent
dynamics involving the gravitational field. As the first step for the
investigation of vacuum densities we evaluate the Wightman function. This
function gives comprehensive insight into vacuum fluctuations and determines
the response of a particle detector of the Unruh-DeWitt type. The problem
under consideration is also of separate interest as an example with
gravitational and topologically-induced polarizations of the vacuum, where
all calculations can be performed in a closed form.

The paper is organized as follows. In Section \ref{sec2} we present the
background associated with the geometry under consideration and the solution
of the Klein-Gordon equation by using \textit{Poincar\'{e}} coordinates and
admitting an arbitrary curvature coupling. Also we present an integral
representation of the Wightman function for an arbitrary planar angle
deficit. In Sections \ref{sec3} and \ref{sec4} we evaluate the parts in the
VEVs of the field squared and the energy-momentum tensor induced by the
cosmic string. Finally, the main results are summarized in Section \ref{conc}%
. In this paper we shall use the units $\hbar =G=c=1$.

\section{Wightman function}

\label{sec2}

The main objective of this section is to obtain the positive frequency
Wightman function associated with a massive scalar field in a
higher-dimensional AdS spacetime in presence of a cosmic string. This
function is important in the calculation of vacuum polarization effects. In
order to do that we first obtain the complete set of normalized mode
functions for the Klein-Gordon equation admitting an arbitrary curvature
coupling.

In cylindrical coordinates, the geometry associated with a cosmic string in
a 4-dimensional AdS spacetime is given by the line element below
(considering a static string along the $y$-axis):
\begin{equation}
ds^{2}=e^{-2y/a}(-dt^{2}+dr^{2}+r^{2}d\phi ^{2})+dy^{2}\ ,  \label{ds1}
\end{equation}%
where $r\geqslant 0$ and $\phi \in \lbrack 0,\ 2\pi /q]$ define the
coordinates on the conical geometry, $(t,\ y)\in (-\infty ,\ \infty )$, and
the parameter $a$ determines the curvature scale of the background
spacetime. The parameter $q$ is bigger than unity and codifies the presence
of the cosmic string. By using \textit{Poincar\'{e}} coordinate defined by $%
z=ae^{y/a}$, the line element above is written in the form conformally
related to the line element associated with a cosmic string in Minkowski
spacetime:
\begin{equation}
ds^{2}=(a/z)^{2}(-dt^{2}+dr^{2}+r^{2}d\phi ^{2}+dz^{2})\ .  \label{ds2}
\end{equation}%
For the new coordinate one has $z\in \lbrack 0,\ \infty )$. Limiting values $%
z=0$ and $z=\infty $ correspond to the AdS boundary and horizon,
respectively.

For an infinite straight cosmic string in the background of Minkowski
spacetime the line element (expression inside the brackets of the right-hand
side of (\ref{ds2})) has been derived in \cite{Vile81} by making use of two
approximations: the weak-field approximation and the thin-string one. In
this case the parameter $q$ is related to the mass per unit length $\mu $ of
the string by the formula $1/q=1-4G\mu $, where $G$ is the Newton's
gravitational constant. In the standard scenario of the cosmic string
formation in the early universe one has $G\mu \sim (\eta /m_{\mathrm{Pl}%
})^{2}$, where $\eta $ is the energy scale of the phase transition at which
the string is formed and $m_{\mathrm{Pl}}$ is the Planck mass. For GUT scale
strings $G\mu \ll 1$ and the weak-field approximation is well justified. The
validity of the line element with the planar angle deficit has been extended
beyond linear perturbation theory by several authors \cite{Gott85} (see also
\cite{V-S}). In this case the parameter $q$ need not to be close to 1. An
interesting limiting case with $q\gg 1$ has been discussed in \cite{V-S}.
Note that in braneworld scenarios based on AdS spacetime, to which the
results given below in the present paper could be applied, the fundamental
Planck scale is much smaller than $m_{\mathrm{Pl}}$ and can be of the order
of string formation energy scale. Similar to the string in Minkowski
spacetime, the line element (\ref{ds1}) is an exact solution of the Einstein
equation in the presence of negative cosmological constant and the string
\cite{Ghe1,Cristine}, for arbitrary value of $q$.

Note that conical defects with the parameter $q$ essentially different from
unity appear also in a number of condensed matter systems. An example is the
graphitic cone, the long-wavelength electronic properties of which are well
described by a Dirac-like model for electronic states in graphene. Graphitic
cones are obtained from the graphene sheet if one or more sectors are
excised. The opening angle of the cone is related to the number of sectors
removed, $N$, by the formula $2\pi (1-N/6)$, with $N=1,2,\ldots ,5$. All
these angles have been observed in experiments \cite{Kris97}.

The generalization of (\ref{ds2}) to $D$-dimensional AdS spacetimes is done
in the usual way, by adding extra Euclidean coordinates:
\begin{equation}
ds^{2}=(a/z)^{2}(-dt^{2}+dr^{2}+r^{2}d\phi
^{2}+dz^{2}+\sum_{i=1}^{d}dz_{i}^{2})\ ,  \label{ds3}
\end{equation}%
where $d=D-4$. Note that the curvature scale $a$ is related to the
cosmological constant, $\Lambda $, and the Ricci scalar, $R$, by the
formulas
\begin{equation}
\Lambda =-\frac{(D-1)(D-2)}{2a^{2}},\ \ R=-\frac{D(D-1)}{a^{2}}\ .
\label{LamR}
\end{equation}

The field equation that will be considered is
\begin{equation}
(\nabla _{\mu }\nabla ^{\mu }-m^{2}-\xi R)\Phi (x)=0\ ,  \label{KG}
\end{equation}%
where $\xi $ is a curvature coupling constant. In the coordinate system
defined by (\ref{ds3}), the complete set of solutions to this equation,
regular at the boundary $z=0$, is characterized by the set of quantum number
$\sigma =(\lambda ,\ n,\ p,\mathbf{k})$ and is given by:
\begin{equation}
\Phi _{\sigma }(x)=\sqrt{\frac{qa^{2-D}p\lambda }{2(2\pi )^{D-3}E}}\ \
z^{(D-1)/2}J_{\nu }(\lambda z)J_{|n|q}(pr)e^{i(nq\phi +\mathbf{k}\cdot
\mathbf{x}-Et)}\ ,  \label{sol1}
\end{equation}%
with $\mathbf{x}=(z,z_{1},\ldots ,z_{d})$,
\begin{equation}
(\lambda ,\ p)\in \lbrack 0,\infty )\ ,\ k_{i}\in (-\infty ,\ \infty ),\
n=0,\pm 1,\pm 2,\ ...  \label{Ranges}
\end{equation}%
In (\ref{sol1}), $J_{\mu }(u)$ represents the Bessel function, $E=\sqrt{%
\lambda ^{2}+p^{2}+\mathbf{k}^{2}}$, and
\begin{equation}
\nu =\sqrt{(D-1)^{2}/4-\xi D(D-1)+m^{2}a^{2}}.  \label{nu}
\end{equation}%
The mode functions in (\ref{sol1}) are normalized by the standard
Klein-Gordon orthonormalization condition
\begin{equation}
i\int d^{D-1}x\sqrt{|g|}g^{00}[\Phi _{\sigma }(x)\partial _{t}\Phi _{\sigma
^{\prime }}^{\ast }(x)-\Phi _{\sigma ^{\prime }}^{\ast }(x)\partial _{t}\Phi
_{\sigma }^{\ast }(x)]=\delta _{\sigma ,\sigma ^{\prime }}\ ,
\label{normcond}
\end{equation}%
where the integral is evaluated over the spatial hypersurface $t=\mathrm{%
const}$, and $\delta _{\sigma ,\sigma ^{\prime }}$ represents the
Kronecker-delta for discrete indices and the Dirac delta function for
continuous ones.

We now employ the mode-sum formula to evaluate the positive frequency
Wightman function:
\begin{equation}
G(x,x^{\prime })=\sum_{\sigma }\Phi _{\sigma }(x)\Phi _{\sigma }^{\ast
}(x^{\prime })\ .  \label{Green}
\end{equation}%
Substituting (\ref{sol1}) into (\ref{Green}) we obtain
\begin{eqnarray}
G(x,x^{\prime }) &=&\frac{q\ (zz^{\prime })^{(D-1)/2}}{2(2\pi )^{D-3}\
a^{D-2}}\sum_{n=-\infty }^{\infty }e^{inq\Delta \phi }\int d\mathbf{k}%
\int_{0}^{\infty }dp\ p\int_{0}^{\infty }d\lambda \ \frac{\lambda }{E}
\notag \\
&\times &e^{i\mathbf{k}\cdot {\Delta }\mathbf{x}-iE\Delta
t}J_{|n|q}(pr)J_{|n|q}(pr^{\prime })J_{\nu }(\lambda z)J_{\nu }(\lambda
z^{\prime })\ ,  \label{Green1}
\end{eqnarray}%
where $\Delta t=t-t^{\prime }$ and ${\Delta }\mathbf{x}=\mathbf{x}-\mathbf{x}%
^{\prime }$. In order to provide a more workable expression for the Wightman
function, after a Wick rotation, $i\Delta t=\Delta \tau $, we use the
identity:
\begin{equation}
\frac{e^{-\sqrt{\lambda ^{2}+p^{2}+\mathbf{k}^{2}}\Delta \tau }}{\sqrt{%
\lambda ^{2}+p^{2}+\mathbf{k}^{2}}}=\frac{2}{\sqrt{\pi }}\int_{0}^{\infty
}ds\ e^{-s^2(\lambda ^{2}+p^{2}+\mathbf{k}^{2})-\Delta \tau ^{2}/4s^{2}}\ .
\label{Ident}
\end{equation}

Now with the help of \cite{Grad} and after some intermediate steps, we can
express the Wightman function in an integral representation as shown below:
\begin{equation}
G(x,x^{\prime })=\frac{q}{a^{D-2}}\left( \frac{zz^{\prime }}{4\pi }\right)
^{(D-1)/2}\int_{0}^{\infty }\frac{ds}{s^{D}}\ e^{-({\mathcal{V}}%
^{2}/4s^{2})}S_{q}\left( rr^{\prime }/{(2s^{2})}\right) I_{\nu }\left(
zz^{\prime }/{(2s^{2})}\right) \ ,  \label{Green2}
\end{equation}%
where $I_{\nu }\left( u\right) $ is the modified Bessel function,
\begin{equation}
S_{q}(u)=\sum_{n=-\infty }^{\infty }e^{inq\Delta \phi }I_{|n|q}(u)=2%
\sideset{}{'}{\sum}_{n=0}^{\infty }\cos (nq\Delta \phi )I_{nq}(u)
\label{sum}
\end{equation}%
and
\begin{equation}
{\mathcal{V}}^{2}=r^{2}+r^{\prime }{}^{2}+z^{2}+z^{\prime }{}^{2}+{\Delta }%
\mathbf{x}^{2}-\Delta t^{2}\ .  \label{Vcal}
\end{equation}%
In (\ref{sum}) the prime on the sign of summation means that the term $n=0$
should be halved.

In \cite{Aram} (see also \cite{Saha11CP}), a general expression for the
summation (\ref{sum}) is presented in terms of an integral representation:
\begin{eqnarray}
S_{q}(u) &=&\frac{1}{q}\sum_{k}e^{u\cos (2\pi k/q+\Delta \phi )}-\frac{1}{%
2\pi }\sum_{j=\pm 1}\sin (q\pi +jq\Delta \phi )  \notag  \label{Sgeneral} \\
&&\times \int_{0}^{\infty }dx\frac{e^{-u\cosh x}}{\cosh (qx)-\cos (q\pi
+jq\Delta \phi )}\ ,  \label{sum1}
\end{eqnarray}%
where $k$ is an integer number defined by%
\begin{equation}
-q/2+q\Delta \phi /(2\pi )\leqslant k\leqslant q/2+q\Delta \phi /(2\pi ).
\label{krange}
\end{equation}%
We can see that, for integer values of $q$, formula (\ref{sum1}) reduces to
the well-known result \cite{Prud,Spin}:
\begin{equation}
2\sideset{}{'}{\sum}_{n=0}^{\infty }\cos (nq\Delta \phi )I_{nq}\left(
u\right) =\frac{1}{q}\sum_{k=0}^{q-1}e^{u\cos (2k\pi /q-\Delta \phi )}.
\label{SumFormSp}
\end{equation}

Substituting (\ref{sum1}) into (\ref{Green2}), with the help of \cite{Grad}
and after some steps, we obtain:
\begin{eqnarray}
G(x,x^{\prime }) &=&\frac{a^{2-D}\ }{(2\pi )^{D/2}}\left[ \sum_{k}F_{\nu
}(u_{k})-\frac{q}{2\pi }\sum_{j=\pm 1}\sin (q\pi +jq\Delta \phi )\right.
\notag \\
&&\times \left. \int_{0}^{\infty }\ dx\ \frac{F_{\nu }(u_{x})}{\cosh
(qx)-\cos (q\pi +jq\Delta \phi )}\right] \ ,  \label{Green3}
\end{eqnarray}%
where the summation over $k$ goes in accordance with (\ref{krange}) and we
have introduced a new function
\begin{equation}
F_{\nu }(u)=\frac{e^{-i\pi (D/2-1)}Q_{\nu -1/2}^{D/2-1}(u)}{%
(u^{2}-1)^{(D-2)/4}}\ .  \label{Fnu}
\end{equation}%
Here, $Q_{\mu }^{\nu }(u)$ is the associated Legendre function, whose
arguments in (\ref{Green3}) are given by
\begin{eqnarray}
u_{k} &=&1+\frac{r^{2}+r^{\prime }{}^{2}-2rr^{\prime }\cos (\Delta \phi
-2\pi k/q)+(\Delta z)^{2}+(\Delta \mathbf{x})^{2}-\Delta t^{2}}{2zz^{\prime }%
},  \notag \\
u_{x} &=&1+\frac{r^{2}+r^{\prime }{}^{2}+2rr^{\prime }\cosh x+(\Delta
z)^{2}+(\Delta \mathbf{x})^{2}-\Delta t^{2}}{2zz^{\prime }}\ .  \label{ux}
\end{eqnarray}%
The component $k=0$ of the first term of (\ref{Green3}) coincides with the
Wightmann function in a pure AdS spacetime, $G_{\text{AdS}}(x,x^{\prime })$.
Because the analysis of vacuum polarization effects in AdS spacetime in the
absence of the string has been developed in the literature by many authors,
here we are mostly interested with the vacuum quantum effects induced by the
cosmic string. Writing the above Wightman function in the decomposed form
\begin{equation}
G(x,x^{\prime })=G_{\text{AdS}}(x,x^{\prime })+G_{c}(x,x^{\prime })\ ,
\label{Gdec}
\end{equation}%
we shall consider in our future analysis the function $G_{c}(x,x^{\prime })$%
, which corresponds to the correction in the Wightman function introduced by
the cosmic string. Moreover, because the presence of the string does not
modify the local geometry for $r\neq 0$, this component of the Wigthmann
function is finite at the coincidence limit.

We can express the function $F_{\nu }(z)$ in (\ref{Fnu}) in terms of the
hypergeometric function as:%
\begin{equation}
F_{\nu }(u)=\frac{B_{\nu }}{u^{\beta _{\nu }}}F(\beta _{\nu }/2+1/2,\beta
_{\nu }/2;\nu +1;u^{-2}),  \label{FnuHyp}
\end{equation}%
where we use the notations%
\begin{eqnarray}
\beta _{\nu } &=&\nu +(D-1)/2,  \notag \\
B_{\nu } &=&\frac{\sqrt{\pi }\Gamma (\beta _{\nu })}{2^{\nu +1/2}\Gamma (\nu
+1)}.  \label{Benu}
\end{eqnarray}%
In the discussion below, we shall need the asymptotics of the function $%
F_{\nu }(u)$ for large values of the argument and for $u$ close to 1. For
large values of $u$ one has%
\begin{equation}
F_{\nu }(u)\approx B_{\nu }/u^{\beta _{\nu }}.  \label{Fnularge}
\end{equation}%
For $u$ close to 1 we use the linear transformation formula 15.3.6 from \cite%
{Abra72}. In the leading order this gives:%
\begin{equation}
F_{\nu }(u)\approx \frac{\Gamma (D/2-1)}{2(u-1)^{D/2-1}}.  \label{FnuNear1}
\end{equation}

For a conformally coupled massless scalar field we have $\nu =1/2$. By using
the formula 15.1.10 from \cite{Abra72} for the hypergeometric function in (%
\ref{FnuHyp}), we find%
\begin{equation}
F_{\nu }(u)=-\frac{1}{2}\Gamma (D/2-1)\left[ (u+1)^{1-D/2}-(u-1)^{1-D/2}%
\right] .  \label{F12}
\end{equation}%
The corresponding Wightman function is presented in the form%
\begin{eqnarray}
G(x,x^{\prime }) &=&\frac{\Gamma (D/2-1)\ (zz^{\prime })^{D/2-1}}{4\pi
^{D/2}a^{D-2}}\left[ \sum_{k}(u_{k}^{(-)1-D/2}-u_{k}^{(+)1-D/2})\right.
\notag \\
&&\left. -\frac{q}{2\pi }\sum_{j=\pm 1}\sin (q\pi +jq\Delta \phi
)\int_{0}^{\infty }\ dx\ \frac{u_{x}^{(-)1-D/2}-u_{x}^{(+)1-D/2}}{\cosh
(qx)-\cos (q\pi +jq\Delta \phi )}\right] \ ,  \label{WFConf}
\end{eqnarray}%
with the notations%
\begin{eqnarray}
u_{k}^{(\pm )} &=&r^{2}+r^{\prime }{}^{2}-2rr^{\prime }\cos (\Delta \phi
-2\pi k/q)+\left( z\pm z^{\prime }\right) ^{2}{}+(\Delta \mathbf{x}%
)^{2}-\Delta t^{2},  \notag \\
u_{x}^{(\pm )} &=&r^{2}+r^{\prime }{}^{2}+2rr^{\prime }\cosh x+\left( z\pm
z^{\prime }\right) ^{2}+(\Delta \mathbf{x})^{2}-\Delta t^{2}.  \label{uxpm}
\end{eqnarray}%
We could obtain expression (\ref{WFConf}) by using the conformal relation of
the problem under consideration with the problem of a cosmic string in
background of the Minkowski spacetime with a Dirichlet boundary
perpendicular to the string and located at $z=0$ (vacuum polarization
effects in the latter geometry have been investigated in \cite{Aram2} for a
massive scalar field with general curvature coupling parameter). The
presence of the boundary in the Minkowskian counterpart is related to the
fact that for the geometry of a string in AdS spacetime we have chosen
regular mode functions (\ref{sol1}). For a conformally coupled massless
field these functions are conformally related to the mode functions for the
string in Minkowski spacetime which obey the Dirichlet boundary condition at
$z=0$.

\section{Calculation of $\langle \protect\phi ^{2}\rangle $}

\label{sec3}

In this section and in the following, we shall investigate the vacuum
polarization effects induced by the cosmic string in AdS spacetime. Two main
calculations will be performed. The evaluation of the VEV of the field
squared, in the first place, followed by the evaluation of the VEV of the
energy-momentum tensor.

Formally the evaluation of the VEV of the field squared is given taking the
coincidence limit of the arguments in the expression for the Wightman
function:
\begin{equation}
\langle \phi ^{2}\rangle =\lim_{x^{\prime }\rightarrow x}G(x,x^{\prime })\ .
\label{phi21}
\end{equation}%
The coincidence limit in the right-hand side is divergent and some
renormalization procedure is needed. By taking into account the
decomposition of the Wightman function, given by (\ref{Gdec}), we
can see that for $r\neq 0$ the divergences come from the pure AdS
part. The part in the Wightman function induced by the presence of
the cosmic string is finite in the coincidence limit. Of course,
we could expect this result from general arguments. The
divergences in the expectation values of local physical
observables, like field squared and energy-momentum tensor, are
entirely geometrical (for a general discussion see \cite{Birr82}).
For expectation values at a given spacetime point they depend on
the curvature tensor and its contractions at the same point. The
presence of the cosmic string does not change the local geometry
of AdS spacetime for points away from the string and,
consequently, the structure of the divergences remains the same as
in pure AdS spacetime. In particular, they do not depend on the
parameters of the cosmic string. In accordance with the standard
procedure, for the renormalization of the VEV in (\ref{phi21}) we
subtract from the Wightman function in the right-hand side the
corresponding DeWitt-Schwinger expansion truncated at the adiabatic order $%
D+1$. The subtracted terms are determined by the local geometry and they are
not affected by the presence of the string. After the subtraction, the
coincidence limit is finite. Of course, we can still add finite
renormalization terms. But, again, these terms are determined entirely by
the local geometry of the spacetime. They do not depend on the planar angle
deficit and renormalize the pure AdS part only. The topological parts in the
VEV induced by the string, which are the main subject of the present
research, are not touched by both the infinite and finite renormalization
procedures.

As we have discussed above, the VEV of the field squared given in (\ref%
{phi21}), is presented in a decomposed form:
\begin{equation}
\langle \phi ^{2}\rangle =\langle \phi ^{2}\rangle _{\text{AdS}}+\langle
\phi ^{2}\rangle _{c}\ ,  \label{phi23}
\end{equation}%
where $\langle \phi ^{2}\rangle _{\text{AdS}}$ is the VEV in AdS spacetime
in the absence of the cosmic string and $\langle \phi ^{2}\rangle _{c}$ is
the string-induced part. Moreover, the renormalization procedure is needed
only for the first contribution in the right hand-side of (\ref{phi23}). The
second contribution is finite at the coincidence limit for $r\neq 0$. Due to
the maximal symmetry of the AdS spacetime and the vacuum state under
consideration, the renormalized VEV of the field squared, $\langle \phi
^{2}\rangle _{\text{AdS}}$, does not present dependence on the specific
point of the spacetime (for the investigation of the VEVs\ in AdS spacetime
see \cite{Burg85}-\cite{Cald99}). So, as we shall see, the contribution
induced by the string, $\langle \phi ^{2}\rangle _{c}$, will become more
important than $\langle \phi ^{2}\rangle _{\text{AdS}}$ for points near the
string or near the AdS horizon.

According to the discussion of the previous section, here we shall analyze
the VEV of the field squared induced by the cosmic string only. This
contribution is directly obtained from (\ref{Green3}), omitting the $k=0$
term and taking the coincidence limit. Note that in the coincidence limit, $%
u_{k}$ in (\ref{Green3}) is an even function of $k$ and the summation over $%
k $ in the range $-q/2\leqslant k\leqslant q/2$ can be transformed into the
summation over the positive values only. Changing also the integration
variable, the string-induced part is presented in the form
\begin{equation}
\langle \phi ^{2}\rangle _{c}=\frac{2a^{2-D}\ }{(2\pi )^{D/2}}\left[
\sum_{k=1}^{[q/2]}F_{\nu }(w_{k})-\frac{q}{\pi }\int_{0}^{\infty }\ dx\
\frac{\sin (q\pi )F_{\nu }(w_{x})}{\cosh (2qx)-\cos (q\pi )}\right] \ ,
\label{Phi}
\end{equation}%
where $[q/2]$ is the integer part of $q/2$, and
\begin{equation}
w_{k}=1+2\rho ^{2}s_{k}^{2},\quad w_{x}=1+2\rho ^{2}\cosh ^{2}x,\
\label{zkzx}
\end{equation}%
with
\begin{equation}
\rho =r/z\ ,\quad s_{k}=\sin (\pi k/q)\ .  \label{rosk}
\end{equation}%
We observe that the string-induced part is a function of the ratio $r/z$
which is the proper distance from the string, $ar/z$, measured in units of
the AdS curvature radius $a$. This property is a consequence of the maximal
symmetry of AdS spacetime.

For a conformally coupled massless scalar field, by taking into account the
expression (\ref{WFConf}) for the Wightman function, one gets%
\begin{equation}
\langle \phi ^{2}\rangle _{c}=\left( z/a\right) ^{D-2}\left[ \langle \phi
^{2}\rangle _{c}^{\text{(M)}}+\langle \phi ^{2}\rangle _{c,b}^{\text{(M)}}%
\right] ,  \label{phi2cConf}
\end{equation}%
where%
\begin{equation}
\langle \phi ^{2}\rangle _{c}^{\text{(M)}}=\frac{2\Gamma (D/2-1)\ }{(4\pi
)^{D/2}r^{D-2}}g_{D-2}(q),  \label{phi2cM}
\end{equation}%
is the VEV for the boundary-free string geometry in background of flat
spacetime and
\begin{equation}
g_{n}(q)=\sum_{k=1}^{[q/2]}s_{k}^{-n}-\frac{q}{\pi }\int_{0}^{\infty }\ dx\
\frac{\sin (q\pi )\cosh ^{-n}x}{\cosh (2qx)-\cos (q\pi )}.  \label{gnq}
\end{equation}%
In (\ref{phi2cConf}), the term%
\begin{eqnarray}
\langle \phi ^{2}\rangle _{c,b}^{\text{(M)}} &=&-\frac{2\Gamma (D/2-1)\ }{%
(4\pi )^{D/2}}\left[ \sum_{k=1}^{[q/2]}\left( r^{2}s_{k}^{2}+z^{2}\right)
^{1-D/2}\right.  \notag \\
&&\left. -\frac{q}{\pi }\sin (q\pi )\int_{0}^{\infty }\ dx\ \frac{\left(
r^{2}\cosh ^{2}x+z^{2}\right) ^{1-D/2}}{\cosh (2qx)-\cos (q\pi )}\right] ,
\label{phi2cMb}
\end{eqnarray}%
is the part induced by a flat boundary located at $z=0$ in the latter
geometry. Closed expressions for $g_{n}(q)$ can be given for even values of $%
n$. In particular, one has
\begin{equation}
g_{2}(q)=\frac{q^{2}-1}{6},\;g_{4}(q)=\frac{\left( q^{2}-1\right) \left(
q^{2}+11\right) }{90}.  \label{g24}
\end{equation}%
For general $q>1$ the function $g_{n}(q)$ is positive. In \cite{Aram2} we
have evaluated the VEV of the field squared in a $D$-dimensional cosmic
string spacetime, associated with scalar quantum field obeying Dirichlet or
Neumann boundary conditions on a flat surface orthogonal to the string. The
result found for this VEV induced by the string, considering the Dirichlet
boundary and massless field, is conformally related to $\langle \phi
^{2}\rangle _{c}$ with the conformal factor $(z/a)^{D-2}$ (see (\ref%
{phi2cConf})).

Let us discuss the asymptotics of the VEV of the field squared near the
string and at large distances. For points near the string, $\rho \ll 1$, by
using the asymptotic expression (\ref{FnuNear1}), in the leading order we
find%
\begin{equation}
\langle \phi ^{2}\rangle _{c}\approx \frac{2\ \Gamma (D/2-1)}{(4\pi
)^{D/2}(ar/z)^{D-2}}g_{D-2}(q)\ .  \label{phi2near}
\end{equation}%
Comparing with (\ref{phi2cM}), we see that the expression on the right of (%
\ref{phi2near}) coincides with the corresponding expression for the VEV of
the field squared in the geometry of a string in background of Minkowski
spacetime, where the distance from the string is replaced by the proper
distance $ar/z$. As the pure AdS part in the VEV, $\langle \phi ^{2}\rangle
_{\text{AdS}}$, is a constant, we conclude that near the string the
topological part dominates in the total VEV. For a fixed value of the radial
coordinate $r$, the limit under consideration corresponds to large values of
$z$, i.e., to points near the AdS horizon. Hence, we conclude that near the
horizon the string-induced part behaves as $z^{D-2}$.

At large distances from the string, $\rho \gg 1$, we use the asymptotic
expression (\ref{Fnularge}). In the leading order this gives:%
\begin{equation}
\langle \phi ^{2}\rangle _{c}\approx \frac{a^{2-D}\ B_{\nu }g_{2\beta _{\nu
}}(q)}{2^{\beta _{\nu }-1}(2\pi )^{D/2}(r/z)^{2\nu +D-1}}\ .
\label{phi2large}
\end{equation}%
At large distances the total VEV is dominated by the pure AdS part. As it is
seen, the decay of the topological part of the field squared with the
distance is power-law for both massless and massive fields. Note that for a
string in background of the Minkowski spacetime the decay of the VEV\ at
large distances is power-law for a massless field and exponential for a
massive field. Hence, we see that the curvature of the background spacetime
has an essential influence on the VEVs at distances larger than the
curvature radius of the background spacetime. From (\ref{phi2large}) it
follows that, for fixed values of $r$, the string-induced part vanishes on
the AdS boundary as $z^{2\nu +D-1}$.

It can be seen that for integer value of $q$, the expression (\ref{Phi})
takes a simpler form:
\begin{equation}
\langle \phi ^{2}\rangle _{c}=\frac{a^{2-D}\ }{(2\pi )^{D/2}}%
\sum_{k=1}^{q-1}F_{\nu }(w_{k})\ .  \label{phi2qint}
\end{equation}%
Note that, for even values of $q$ the integral term in (\ref{Phi})
contributes as well. Indeed, we assume that $q=2n+\epsilon $ with $\epsilon
\rightarrow 0$. As the integral is divergent at the lower limit, the
dominant contribution for small $\epsilon $ comes from the region near this
limit (small values of $x$). Expanding the integrand, we can see that the
second term in the square brackets of (\ref{Phi}) gives the contribution $%
(1/2)F_{\nu }(1+2\rho ^{2})$. In particular, for a conformally coupled
massless field the formula (\ref{phi2qint}) reduces to%
\begin{equation}
\langle \phi ^{2}\rangle _{c}=\frac{2\Gamma (D/2-1)\ }{(4\pi )^{D/2}a^{D-2}}%
\sum_{k=1}^{q-1}\left[ \frac{1}{(\rho s_{k})^{D-2}}-\frac{1}{\left( 1+\rho
^{2}s_{k}^{2}\right) ^{D/2-1}}\right] .  \label{phi2qintc}
\end{equation}%
For even values of $D$, the sum with the first term in the square brackets
of (\ref{phi2qintc}) can be further simplified by using the recurrence
relation given in \cite{Aram1} and the result $%
\sum_{k=1}^{q-1}s_{k}^{-2}=(q^{2}-1)/3$. So, in the 4-dimensional spacetime
we may write
\begin{equation}
\langle \phi ^{2}\rangle _{c}=\frac{(z/a)^{2}}{16\pi ^{2}}\left( \frac{%
q^{2}-1}{3r^{2}}-\sum_{k=1}^{q-1}\frac{1}{z^{2}+r^{2}s_{k}^{2}}\right) \ .
\label{phi2D4}
\end{equation}%
For a six-dimensional spacetime, the summation needed is: $%
\sum_{k=1}^{q-1}\sin ^{-4}(\pi k/q)=(q^{2}-1)(q^{2}+11)/45$. As a result,
for the VEV we obtain:
\begin{equation}
\langle \phi ^{2}\rangle _{c}=\frac{(z/a)^{4}}{64\pi ^{3}}\left[ \frac{%
(q^{2}-1)(q^{2}+11)}{45r^{4}}-\sum_{k=1}^{q-1}\frac{1}{%
(z^{2}+r^{2}s_{k}^{2})^{2}}\right] \ .  \label{phi2qint6}
\end{equation}

\begin{figure}[tbph]
\begin{center}
\epsfig{figure=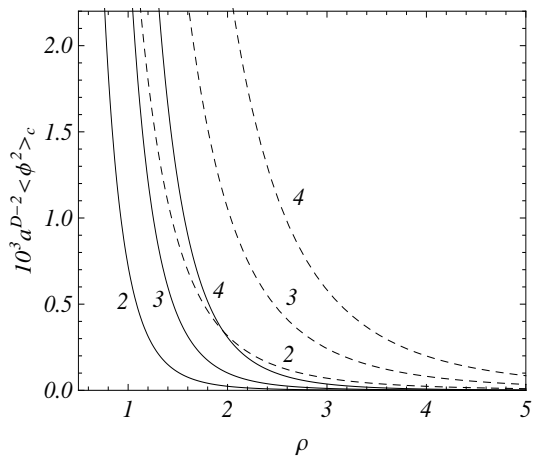,width=7.cm,height=6.cm}
\end{center}
\caption{The string-induced part in the VEV of the field squared as a
function of the proper distance from the string for $D=4$ minimally (full
curves) and conformally (dashed curves) massless scalar fields. The numbers
near the curves correspond to the values of the parameter $q$.}
\label{fig1}
\end{figure}

In figure \ref{fig1} we have displayed the dependence of the string-induced
part in the VEV of the field squared versus the proper distance from the
string (measured in units of the AdS curvature radius). The graphs are
plotted for $D=4$ minimally (full curves) and conformally (dashed curves)
coupled massless scalar fields for separate values of the parameter $q$
(numbers near the curves).

\section{VEV of the energy-momentum tensor}

\label{sec4}

Similar to the case of the field squared, the VEV\ of the energy-momentum
tensor is presented in the decomposed form:
\begin{equation}
\langle T_{\mu \nu }\rangle =\langle T_{\mu \nu }\rangle _{\text{AdS}%
}+\langle T_{\mu \nu }\rangle _{c}\ ,  \label{TmunuDec}
\end{equation}%
where the first and second terms on the right-hand side correspond to the
purely AdS and string-induced parts, respectively. Because of the maximal
symmetry of AdS spacetime and the vacuum state under consideration, the pure
AdS part in the VEV\ of the energy-momentum tensor is proportional to the
metric tensor: $\langle T_{\mu \nu }\rangle _{AdS}=g_{\mu \nu }\langle
T_{\beta }^{\beta }\rangle _{\text{AdS}}/D$, with $\langle T_{\beta }^{\beta
}\rangle _{AdS}$ being the corresponding trace (the expression for the
latter in arbitrary number of spacetime dimensions can be found in \cite%
{Cald99}). In this section we shall analyze the contribution induced by the
string. To develop this calculation we shall use the following expression:
\begin{equation}
\langle T_{\mu \nu }\rangle _{c}=\lim_{x^{\prime }\rightarrow x}\partial
_{\mu ^{\prime }}\partial _{\nu }G_{c}(x,x^{\prime })+\left[ \left( \xi -{1}/%
{4}\right) g_{\mu \nu }\nabla _{\alpha }\nabla ^{\alpha }-\xi \nabla _{\mu
}\nabla _{\nu }+\xi R_{\mu \nu }\right] \langle \phi ^{2}\rangle _{c}\ ,
\label{EMT}
\end{equation}%
where for the AdS spacetime, the Ricci tensor reads
\begin{equation}
R_{\mu \nu }=-(D-1)g_{\mu \nu }a^{-2}\ .  \label{Rmunu}
\end{equation}

We may write the covariant d'Alembertian of $\langle \phi ^{2}\rangle _{c}$,
appearing in (\ref{EMT}), as shown below:
\begin{equation}
\nabla _{\alpha }\nabla ^{\alpha }\langle \phi ^{2}\rangle _{c}=\frac{%
8a^{-D}\ }{(2\pi )^{D/2}}\left[ \sum_{k=1}^{[q/2]}f(w_{k},s_{k})-\frac{q}{%
\pi }\int_{0}^{\infty }dx\frac{\sin (q\pi )f(w_{x},\cosh x)}{\cosh
(2qx)-\cos (q\pi )}\right] \ ,  \label{Dalphi2}
\end{equation}%
with the notation%
\begin{equation}
f(u,v)=(u-1)\left( 2v^{2}+u-1\right) F_{\nu }^{\prime \prime }(u)+[2v^{2}+%
\frac{D+1}{2}(u-1)]F_{\nu }^{\prime }(u),  \label{fuv}
\end{equation}%
and the prime means the derivative of the functions with respect to their
arguments. By using the formula 15.2.3 for the hypergeometric function from
\cite{Abra72}, the derivatives appearing in (\ref{fuv}) are written in the
form%
\begin{eqnarray}
F_{\nu }^{\prime }(u) &=&-\frac{\beta _{\nu }B_{\nu }}{u^{\beta _{\nu }+1}}%
F(\beta _{\nu }/2+1/2,\beta _{\nu }/2+1;c;u^{-2}),  \notag \\
F_{\nu }^{\prime \prime }(u) &=&\frac{\beta _{\nu }\left( \beta _{\nu
}+1\right) B_{\nu }}{u^{\beta _{\nu }+2}}F(\beta _{\nu }/2+3/2,\beta _{\nu
}/2+1;c;u^{-2}).  \label{Fnuder}
\end{eqnarray}

By using the the expression for the Wightman function and the VEV of the
field squared, after a long but straightforward calculations, the diagonal
components of the energy-momentum tensor are presented in the form (no
summation over $\mu $):
\begin{equation}
\langle T_{\mu }^{\mu }\rangle _{c}=\frac{2a^{-D}}{(2\pi )^{D/2}}\left[
\sum_{k=1}^{[q/2]}G_{\mu }(w_{k},s_{k})-\frac{q}{\pi }\int_{0}^{\infty }\
dx\ \frac{\sin (q\pi )G_{\mu }(w_{x},\cosh x)}{\cosh (2qx)-\cos (q\pi )}%
\right] \ ,  \label{T-diag}
\end{equation}%
where $s_{k}$ is given in (\ref{rosk}),
\begin{equation}
G_{\mu }(u,v)=f_{\mu }(u,v)+(4\xi -1)f(u,v)-\xi (D-1)F_{\nu }(u),
\label{Gmu}
\end{equation}%
and
\begin{eqnarray}
f_{0}(u,v) &=&-[1+2\xi (u-1)]F_{\nu }^{\prime }(u)\ ,  \notag \\
f_{1}(u,v) &=&[2v^{2}\left( 1-2\xi \right) -1-2\xi (u-1)]F_{\nu }^{\prime
}(u)+2v^{2}(1-4\xi )(u-1)F_{\nu }^{\prime \prime }(u)\ ,  \notag \\
f_{2}(u,v) &=&-[1+2v^{2}\left( 2\xi -1\right) +2\xi (u-1)]F_{\nu }^{\prime
}(u)-2(u-1)(1-v^{2})F_{\nu }^{\prime \prime }(u)\ ,  \notag \\
f_{3}(u,v) &=&[\left( 1-4\xi \right) (u-1)-1]F_{\nu }^{\prime
}(u)+(u-1)^{2}(1-4\xi )F_{\nu }^{\prime \prime }(u)\ .  \label{fmuuv}
\end{eqnarray}%
For the components $\langle T_{\mu }^{\mu }\rangle _{c}$ with $\mu =4\ ...\
D-1$, one has $\langle T_{\mu }^{\mu }\rangle _{c}=\langle T_{0}^{0}\rangle
_{c}$. The latter relation is a direct consequence of the invariance of the
problem with respect to the boosts along the corresponding directions. The
last two terms in the right-hand side of (\ref{Gmu}) are the same for all
diagonal components and come from the first end last terms in the square
brackets of (\ref{EMT}). Note that the string-induced part in the energy
density is given by $-\langle T_{0}^{0}\rangle _{c}$. For the non-zero
off-diagonal component we have,
\begin{equation}
\langle T_{3}^{1}\rangle _{c}=\frac{2a^{-D}}{(2\pi )^{D/2}\rho }\left[
\sum_{k=1}^{[q/2]}G(w_{k})-\frac{q}{\pi }\int_{0}^{\infty }\ dx\ \frac{\sin
(q\pi )G(w_{x})}{\cosh (2qx)-\cos (q\pi )}\right] \ ,  \label{T31}
\end{equation}%
where
\begin{equation}
G(u)=(u-1)\left[ (u-1)(4\xi -1)F_{\nu }^{\prime \prime }(u)+(2\xi -1)F_{\nu
}^{\prime }(u)\right] \ .  \label{geu}
\end{equation}

It can be explicitly checked that the string-induced parts in the VEVs obey
the trace relation%
\begin{equation}
\langle T_{\mu }^{\mu }\rangle _{c}=D(\xi -\xi _{D})\nabla _{\alpha }\nabla
^{\alpha }\langle \phi ^{2}\rangle _{c}+m^{2}\langle \phi ^{2}\rangle _{c}.
\label{trace}
\end{equation}%
In particular, the topological part in the VEV of the energy-momentum tensor
is traceless for a conformally coupled massless scalar field. The trace
anomaly appears in the pure AdS part only. Note that for a $D=4$ conformally
coupled massless scalar field (see, for example, \cite{Camp92}) $\langle
T_{\mu }^{\mu }\rangle _{\text{AdS}}=-1/(240\pi ^{2}a^{4})$. We have also
observed that for a massless conformally coupled field, the above
expressions are conformally related with the corresponding ones obtained in
\cite{Aram2} with the conformal factor $(z/a)^{D}$. In particular, for the
off-diagonal component we have%
\begin{eqnarray}
\langle T_{3}^{1}\rangle _{c} &=&-\frac{2\Gamma (D/2+1)\rho }{(4\pi
)^{D/2}(D-1)a^{D}}\bigg[\sum_{k=1}^{[q/2]}\frac{s_{k}^{2}}{(1+\rho
^{2}s_{k}^{2})^{D/2+1}}  \notag \\
&&-\frac{q}{\pi }\sin (q\pi )\int_{0}^{\infty }\ dx\ \frac{(1+\rho ^{2}\cosh
^{2}x)^{-D/2-1}}{\cosh (2qx)-\cos (q\pi )}\cosh ^{2}x\bigg].  \label{T31cm0}
\end{eqnarray}%
Note that in this case the off-diagonal component vanishes on the string. As
it will be seen below, the diagonal components diverge on the string.

Due to the presence of the off-diagonal component, from the covariant
conservation equation for the energy-momentum tensor, $\nabla _{\mu }\langle
T_{\nu }^{\mu }\rangle _{c}=0$, two non-trivial differential equations
follow:
\begin{equation}
\frac{1}{r}\partial _{r}\left( r\langle T_{1}^{1}\rangle _{c}\right) -\frac{1%
}{r}\langle T_{2}^{2}\rangle _{c}-\frac{D}{z}\langle T_{1}^{3}\rangle
_{c}+\partial _{z}\langle T_{1}^{3}\rangle _{c}=0  \label{ConsEq1}
\end{equation}%
and
\begin{equation}
\frac{1}{r}\partial _{r}\left( r\langle T_{3}^{1}\rangle _{c}\right) -\frac{D%
}{z}\langle T_{3}^{3}\rangle _{c}+\partial _{z}\langle T_{3}^{3}\rangle _{c}+%
\frac{1}{z}\langle T_{\mu }^{\mu }\rangle _{c}=0\ .  \label{Conseq2}
\end{equation}%
It can be explicitly checked that the above relations are obeyed by the
string-induced parts in the VEV of the energy-momentum tensor, given by
expressions (\ref{T-diag}) and (\ref{T31}).

Relatively simple expressions for the VEV of the energy-momentum tensor are
obtained near the string and at large distances. First we consider the
region near the string, assuming that $\rho \ll 1$. By taking into account (%
\ref{FnuNear1}), in the leading order for the diagonal components we find
(no summation over $\mu $):%
\begin{equation}
\langle T_{\mu }^{\mu }\rangle _{c}\approx -\frac{\Gamma (D/2)}{(4\pi
)^{D/2}(a\rho )^{D}}\left[ f_{\mu }^{(0)}g_{D}(q)+f_{\mu }^{(1)}g_{D-2}(q)%
\right] ,  \label{Tnear}
\end{equation}%
with the notations%
\begin{eqnarray}
f_{\mu }^{(0)} &=&-f_{2}^{(0)}/(D-1)=-1,\quad \mu \neq 2,  \notag \\
f_{0}^{(1)} &=&(D-2)\left( 1-4\xi \right) ,\quad
f_{1}^{(1)}=-f_{2}^{(1)}/(D-1)=4\xi ,  \label{fmu12}
\end{eqnarray}%
and $f_{\mu }^{(j)}=f_{0}^{(j)}$ for $\mu \geqslant 3$, $j=1,2$. The leading
term given by (\ref{Tnear}) coincides with the corresponding expression for
a string in Minkowski spacetime with the distance from the string replaced
by $ar/z$. For the off-diagonal component the leading term is given by the
expression%
\begin{equation}
\langle T_{3}^{1}\rangle _{c}\approx \left( \xi -\xi _{D}\right) \frac{%
4(D-1)\Gamma (D/2)}{(4\pi )^{D/2}a(ar/z)^{D-1}}g_{D-2}(q)\ ,  \label{T31near}
\end{equation}%
with the function $g_{n}(q)$ defined in (\ref{gnq}). Note that the
off-diagonal component is suppressed by an additional factor $r/z$ compared
with the diagonal ones. For a conformally coupled massless field the
behavior of the off-diagonal component near the string directly follows from
(\ref{T31cm0}). In this case the off-diagonal component vanishes on the
string. For $D=4$, in (\ref{Tnear}) and (\ref{T31near}) we can use the
expressions (\ref{g24}). In this case it can be seen that for a conformally
coupled scalar field the energy density near the string is negative. For a
minimally coupled field the energy density is positive for $q^{2}<19$ and
negative for $q^{2}>19$. For fixed values of $r$, the limit under
consideration corresponds to points near the AdS horizon. In particular, we
see that on the horizon the VEVs of the diagonal components of the
energy-momentum tensor diverge as $z^{D}$. The asymptotic expressions show
that near the string or near the horizon the topological part is large and
the back-reaction of the quantum effects on the bulk geometry is important.

Now we turn to the asymptotic of the string-induced part at large distances
from the string, $\rho \gg 1$. By using the asymptotic expression (\ref%
{Fnularge}), to the leading order for the diagonal components we get (no
summation over $\mu $):%
\begin{equation}
\langle T_{\mu }^{\mu }\rangle _{c}\approx -\frac{B_{\nu }\nu \left[ \beta
_{\nu }-2\xi \left( 2\nu +D\right) \right] }{2^{\beta _{\nu }-1}(2\pi
)^{D/2}a^{D}(r/z)^{2\nu +D-1}}g_{2\beta _{\nu }}(q),  \label{Tlarge}
\end{equation}%
for $\mu \neq 3$, and%
\begin{equation}
\langle T_{3}^{3}\rangle _{c}\approx -\frac{D-1}{2\nu }\langle
T_{0}^{0}\rangle _{c}.  \label{T3large}
\end{equation}%
For both minimally and conformally coupled fields the energy density
corresponding to (\ref{Tlarge}) is positive. For the off-diagonal component
one has
\begin{equation}
\langle T_{3}^{1}\rangle _{c}\approx \frac{\beta _{\nu }z}{\nu r}\langle
T_{0}^{0}\rangle _{c}.  \label{T31large}
\end{equation}%
The latter is suppressed by an additional factor $z/r$ compared with the
diagonal components. Similar to the case of the field squared, the decay of
the topological part in the VEV of the energy-momentum tensor is power-law
for both massless and massive fields. Note that, at large distances, the
stresses in the subspace perpendicular to the $z$-axis are isotropic. For a
minimally coupled massless scalar field, the relation (\ref{T3large})
between the energy density and the stress along the $z$-axis is of the
cosmological constant type. Note that, for a fixed value of the radial
coordinate, the limit under consideration corresponds to points near the AdS
boundary. From the asymptotic expressions we see that the string-induced
parts in the diagonal components vanish on the AdS boundary as $z^{2\nu
+D-1} $.

In the case of integer $q$, the general formulas (\ref{T-diag}) and (\ref%
{T31}) take the form (no summation over $\mu $)
\begin{eqnarray}
\langle T_{\mu }^{\mu }\rangle _{c} &=&\frac{\ a^{-D}}{(2\pi )^{D/2}}%
\sum_{k=1}^{q-1}G_{\mu }(w_{k},s_{k})\ ,  \notag \\
\langle T_{3}^{1}\rangle _{c} &=&\frac{a^{-D}}{(2\pi )^{D/2}}%
\sum_{k=1}^{q-1}G(w_{k},s_{k})\ ,  \label{T31Intq}
\end{eqnarray}%
with functions $G_{\mu }(u,v)$ and $G(u,v)$ defined in (\ref{Gmu}) and (\ref%
{geu}). For the general situation, the complete expression for each
component of the VEV of the energy-momentum tensor is long one. However, we
can obtain simpler expressions for $\nu =1/2$, which corresponds to a
conformally coupled massless field. Below we present the expressions for the
topological part in the VEV of the energy-momentum tensor for this case.

First we write down the diagonal components in the 4-dimensional AdS
spacetime (no summation over $\mu $),
\begin{equation}
\langle T_{\mu }^{\mu }\rangle _{c}=\frac{(a\rho )^{-4}}{96\pi ^{2}}%
\sum_{k=1}^{q-1}\frac{f_{4,\mu }(\rho s_{k},s_{k})}{s_{k}^{4}(1+\rho
^{2}s_{k}^{2})^{3}}\ ,  \label{TmucD4}
\end{equation}%
with
\begin{eqnarray}
f_{4,0}(u,v) &=&-u^{6}+\left( 9-8v^{2}\right) u^{4}+\left( 3u^{2}+1\right)
\left( 3-2v^{2}\right) \ ,  \notag \\
f_{4,1}(u,v) &=&-u^{6}+\left( 9-4v^{2}\right) u^{4}+\left( 3u^{2}+1\right)
\left( 3-2v^{2}\right) \ ,  \notag \\
f_{4,2}(u,v) &=&-u^{6}-\left( 27-20v^{2}\right) u^{4}-3\left(
3u^{2}+1\right) \left( 3-2v^{2}\right) \ ,  \notag \\
f_{4,3}(u,v) &=&3u^{6}+\left( 9-8v^{2}\right) u^{4}+\left( 3u^{2}+1\right)
\left( 3-2v^{2}\right) \ ,  \label{fspecial}
\end{eqnarray}%
being $\rho =r/z$. As to the off-diagonal component, we have
\begin{equation}
\langle T_{3}^{1}\rangle _{c}=-\frac{\rho }{24\pi ^{2}a^{4}}\sum_{k=1}^{q-1}%
\frac{s_{k}^{2}}{\left( 1+\rho ^{2}s_{k}^{2}\right) ^{3}}\ .  \label{T31cD4}
\end{equation}

For the 6-dimensional case, the VEV of the energy-momentum tensor reads (no
summation over $\mu $),
\begin{equation}
\langle T_{\mu }^{\mu }\rangle _{c}=\frac{(a\rho )^{-6}}{640\pi ^{3}}%
\sum_{k=1}^{q-1}\frac{f_{6,\mu }(\rho s_{k},s_{k})}{s_{k}^{6}(1+\rho
^{2}s_{k}^{2})^{4}}\ ,  \label{TmucD6}
\end{equation}%
with
\begin{eqnarray}
f_{6,l}(u,v) &=&-u^{8}+2\left( 10-9v^{2}\right)
u^{6}+(6u^{4}+4u^{2}+1)\left( 5-4v^{2}\right) \ ,  \notag \\
f_{6,1}(u,v) &=&-u^{8}+4\left( 5-3v^{2}\right) u^{6}+(6u^{4}+4u^{2}+1)\left(
5-4v^{2}\right) \ ,  \notag \\
f_{6,2}(u,v) &=&-u^{8}-4\left( 25-21v^{2}\right)
u^{6}-5(6u^{4}+4u^{2}+1)\left( 5-4v^{2}\right) \ ,  \notag \\
f_{6,3}(u,v) &=&5u^{8}+2\left( 10-9v^{2}\right)
u^{6}+(6u^{4}+4u^{2}+1)\left( 5-4v^{2}\right) \ ,  \label{f6mu}
\end{eqnarray}%
$l=0,4,5$, for diagonal components, and
\begin{equation}
\langle T_{3}^{1}\rangle _{c}=-\frac{3\rho }{320\pi ^{3}a^{6}}%
\sum_{k=1}^{q-1}\frac{s_{k}^{2}}{\left( 1+\rho ^{2}s_{k}^{2}\right) ^{4}}\ ,
\label{T31cD6}
\end{equation}%
for the off-diagonal component. As it has been mentioned before, all but
off-diagonal component diverge for $\rho \rightarrow 0$.

\begin{figure}[tbph]
\begin{center}
\epsfig{figure=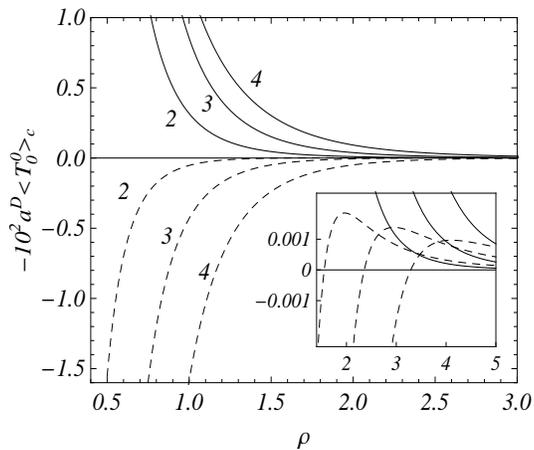,width=7.cm,height=6.cm}
\end{center}
\caption{The string-induced part in the VEV of the energy density as a
function of the proper distance from the string (measured in units of the
AdS curvature radius) for $D=4$ minimally (full curves) and conformally
(dashed curves) massless scalar fields. The numbers near the curves
correspond to the values of the parameter $q$.}
\label{fig2}
\end{figure}

Figure \ref{fig2} presents the dependence of the string-induced part in the
VEV of the energy density as a function of the proper distance from the
string for $D=4$ minimally (full curves) and conformally (dashed curves)
massless scalar fields. The numbers near the curves correspond to the values
of the parameter $q$. The behavior for large values $\rho $ is plotted
separately in the inset to show that for both cases of minimally and
conformally coupled fields the energy density goes to zero being positive.

\begin{figure}[tbph]
\begin{center}
\epsfig{figure=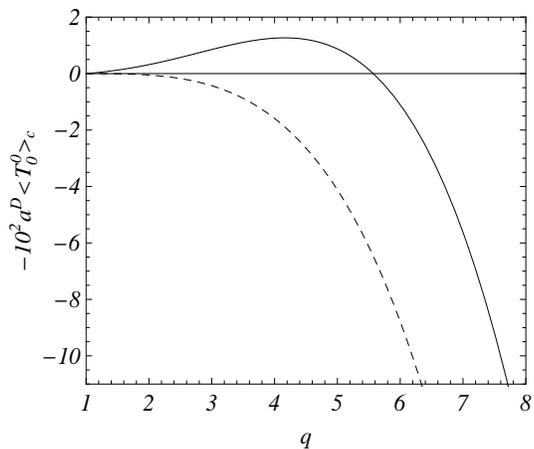,width=7.cm,height=6.cm}
\end{center}
\caption{The string-induced part in the VEV of the energy density as a
function of the parameter $q$ for fixed proper distance from the string
corresponding to $\protect\rho =1$. The full and dashed curves are for
minimally and conformally coupled $D=4$ massless scalar fields.}
\label{fig3}
\end{figure}

In figure \ref{fig3} we plot the string-induced part in the energy density
versus the parameter $q$ for a fixed value of the distance from the string
corresponding to $\rho =1$. The full and dashed curves correspond to
minimally and conformally coupled massless fields in 4-dimensional AdS
spacetime. For the conformal coupling the energy density is negative,
whereas for the minimal coupling it is positive for $q<5.6$ and is negative
for larger values of $q$.

\section{Conclusion}

\label{conc}

In this paper we have calculated the VEV of the field squared and the
energy-momentum tensor associated with a scalar field in a $D$-dimensional
AdS spacetime in the presence of an idealized cosmic string. In fact,
because the calculations of the VEVs in a purely AdS spacetime have been
done by several authors, here we were mainly interested in the topological
parts in the VEVs induced by the presence of the string. These objective
were satisfactorily attained because, for this system, the Wightman function
could be expressed as the sum of two terms: the first one due to the AdS
background itself in the absence of string, and the second one induced by
the presence of the string. This allowed us to write the VEVs as the sum of
two contributions following the same structure of the Wightman function.
Moreover, because the presence of the string does not modify the curvature
of the AdS background, all the divergences presented in the calculations of $%
\langle \phi ^{2}\rangle $ and $\langle T_{\nu }^{\mu }\rangle $, appear
only in the contributions due the purely AdS space. So the contributions
induced by the string do not require renormalization. All of them are
automatically finite for points outside the string. By using (\ref{sum1}),
we were able to express the Wightman function in a compact form, Eq. (\ref%
{Green3}), allowing us to analyze the VEVs of the field squared and the
energy-momentum tensor in a systematic way.

The VEV of the field squared is presented in the decomposed form, Eq. (\ref%
{phi23}), where the part induced by the string is given by the expression (%
\ref{Phi}). For a conformally coupled massless scalar field this expression
is conformally related to the corresponding quantity for the geometry of a
cosmic string in flat spacetime with an additional flat boundary with
Dirichlet boundary condition on it (see (\ref{phi2cConf})). At small proper
distances from the string, compared with the AdS curvature radius, the
leading term coincides with the corresponding expression for the VEV of the
field squared in the geometry of a string in background of flat spacetime,
where the distance from the string is replaced by the proper distance $ar/z$%
. In this limit the VEV behaves as $(ar/z)^{2-D}$. For a fixed value of the
radial coordinate this corresponds to points near the AdS horizon. At large
distances from the string, the topological part in the VEV of the field
squared decays as $(ar/z)^{1-2\nu -D}$ with the parameter $\nu $ defined by (%
\ref{nu}). In this regime the decay with the distance is power-law for both
massless and massive fields. It is worth to note that for a string in
background of flat spacetime the decay of the VEV\ at large distances is
power-law for a massless field and exponential for a massive field. We see
that the curvature of the background spacetime has an essential influence on
the VEVs at distances larger than the curvature radius of the background
spacetime. The expression for the VEV of the field squared takes a simpler
form, Eq. (\ref{phi2qint}), for integer values of the parameter $q$ related
to the angle deficit in the cosmic string geometry. In particular, for a
conformally coupled massless field one has the expression (\ref{phi2qintc}).

Similar to the case of the field squared, the VEV of the energy-momentum
tensor is decomposed as the sum of the pure AdS and string-induced parts.
The presence of the string breaks the maximal symmetry of AdS spacetime and
the vacuum energy-momentum tensor is non-diagonal. The diagonal components
are given by the expressions (\ref{T-diag}) and the non-zero off-diagonal
component is presented by the expression (\ref{T31}). We have explicitly
checked that the topological part in the VEV of the energy-momentum tensor
obeys the trace relation (\ref{trace}) and the covariant conservation
equation. In particular, this part is traceless for a conformally coupled
massless scalar field. The trace anomaly in the total energy-momentum tensor
is contained in the pure AdS part only. Relatively simple expressions for
the VEV of the energy-momentum tensor are obtained near the string and at
large distances. Near the string, the topological parts in the diagonal
components scale as the inverse $D$-th power of the proper distance from the
string, whereas the off-diagonal component scales as the inverse $(D-1)$-th
power of the distance. For a conformally coupled massless scalar field the
off-diagonal component vanishes on the string linearly with the distance.
The pure AdS part in the vacuum energy-momentum tensor is a constant
everywhere and near the string the total VEV is dominated by the topological
part. In this region the back-reaction of the string-induced quantum effects
on the bulk geometry is important. At large distances from the string,
compared with the AdS curvature radius, the decay of the topological part
with the distance is power-law for both massless and massive fields. The
leading terms in the corresponding asymptotic expansions are given by
expressions (\ref{Tlarge})-(\ref{T31large}). These expressions describe also
the behavior of the VEVs for points near the AdS boundary. For a fixed value
of the radial coordinate, the string-induced parts vanish on the AdS
boundary as $z^{2\nu +D-1}$. The general formulas for the VEV of the
energy-momentum tensor are simplified in the special case of integer values
for the parameter $q$. The corresponding expressions are given by Eq. (\ref%
{T31Intq}). These expressions are further simplified for a conformally
coupled massless scalar field. We present explicit formulas for the energy
momentum-tensor in spacetime dimensions $D=4$ and $D=6$.

In a way similar to that described above, we can evaluate one-loop quantum
effects induced by the string in AdS spacetime in the presence of branes.
For the corresponding geometry with two branes in background of pure AdS
spacetieme (employed in Randall-Sundrum type braneworld models) the VEVs of
the field squared and the energy-momentum tensor are investigated in \cite%
{Knap04,Saha05} (see also \cite{Saha06,Saha06b} for the models with compact
internal spaces). In the presence of the cosmic string, the mode functions
in the region between the branes are given by expression (\ref{sol1}), with
the function $J_{\nu }(\lambda z)$ replaced by the linear combination of the
Bessel and Neumann functions with the same argument. The eigenvalues for $%
\lambda $ are determined from the boundary conditions on the branes and they
are expressed in terms of the zeros of the cross-product of the Bessel and
Neumann functions. The mode-sum for the Wightman function contains summation
over these zeros. By applying to this sum the generalized Abel-Plana formula
from \cite{SahaBook}, we can present the VEVs as the sum of boundary-free
AdS parts, investigated in the present paper, and the parts induced by the
branes. The investigation for the latter will be presented elsewhere. An
interesting application of the results presented in this paper would be the
investigation of the corresponding effects in the boundary conformal field
theory by using the AdS/CFT correspondence.

\section*{Acknowledgment}

E.R.B.M. thanks Conselho Nacional de Desenvolvimento Cient\'{\i}fico e Tecnol%
\'{o}gico (CNPq) for partial financial support. A.A.S. was supported by the
Program CAPES-PV.


\begin{thebibliography}{99}
\bibitem{Kibble} T. W. Kibble, J. Phys. A \textbf{9}, 1387 (1976).

\bibitem{V-S} A. Vilenkin and E.P.S. Shellard, \textit{Cosmic Strings and
Other Topological Defects} (Cambridge University Press, Cambridge, England,
1994).

\bibitem{Berezinski} V. Berezinski, B. Hnatyk and A. Vilenkin, Phys. Rev. D
\textbf{64}, 043004 (2001).

\bibitem{Damour} T. Damour and A. Vilenkin, Phys. Rev. Lett. \textbf{85},
3761 (2000).

\bibitem{Bhattacharjee} P. Bhattacharjee and G. Sigl, Phys. Rep. \textbf{327}%
, 109 (2000).

\bibitem{Sarangi} S. Sarangi and S.-H. Henry Tye, Phys. Lett. B \textbf{536}%
, 185 (2002).

\bibitem{Copeland} E. J. Copeland, R. C. Myers and J. Polchinski, J. High
Energy Phys. \textbf{06}, 013 (2004).

\bibitem{Dvali} G. Dvali and A. Vilenkin, J. Cosmol. Astropart. Phys.
\textbf{03}, 010 (2004).

\bibitem{scalar} B. Linet, Phys. Rev. D \textbf{35}, 536 (1987).

\bibitem{scalar1} A. G. Smith, in \textit{Symposium on the Formation and
Evolution of Cosmic String}, edited by G. W. Gibbons, S. W. Hawking and T.
Vachaspati (Cambridge University Press, Cambridge, England, 1989).

\bibitem{scalar2} P. C. Davies and V. Sahni, Class. Quantum Grav. \textbf{5}%
, 1 (1987).

\bibitem{scalar3} T. Souradeep and V. Sahni, Phys. Rev. D \textbf{46}, 1616
(1992).

\bibitem{scalar4} M. E. X. Guimar\~{a}es and B. Linet, Class. Quantum Grav.
\textbf{10}, 1665 (1993).

\bibitem{Aram1} E. R. Bezerra de Mello, V. B. Bezerra, A. A. Saharian and A.
S. Torloyan, Phys. Rev. D \textbf{74}, 025017 (2006).

\bibitem{Site11} Yu. A. Sitenko and N. D. Vlasii, arXiv:1110.0731.

\bibitem{ferm} V. P. Frolov and E. M. Serebriany, Phys. Rev. D \textbf{15},
3779 (1987).

\bibitem{ferm1} B. Linet, J. Math. Phys. \textbf{36}, 3694 (1995).

\bibitem{ferm2} E. S. Moreira Jnr., Nucl. Phys. B \textbf{451}, 365 (1995).

\bibitem{ferm3} V. B. Bezerra and N. R. Khusnutdinov, Class. Quantum Grav.
\textbf{23}, 3449 (2006).

\bibitem{Beze08f} E. R. Bezerra de Mello, V. B. Bezerra, A. A. Saharian and
A. S. Tarloyan, Phys. Rev. D \textbf{78}, 105007 (2008).

\bibitem{StringdS} E. R. Bezerra de Mello and A. A. Saharian, JHEP \textbf{04%
} (2009) 046.

\bibitem{StringdS1} E. R. Bezerra de Mello and A. A. Saharian, JHEP \textbf{%
08} (2010) 038.

\bibitem{Ghe1} M. H. Dehghani, A. M. Ghezelbash and R. B. Mann, Nucl. Phys.
B \textbf{625}, 389 (2002).

\bibitem{Cristine} C. A. Ballon Bayona, C. N. Ferreira and V. J. Vasquez
Otoya, Class. Quantum Grav. \textbf{28}, 015011 (2011).

\bibitem{Ahar00} O. Aharony, S. S. Gubser, J. Maldacena, H. Ooguri and Y.
Oz, Phys. Rep. \textbf{323}, 183 (2000).

\bibitem{Brax03} P. Brax and C. Van de Bruck, Classical Quantum Gravity
\textbf{20}, R201(2003).

\bibitem{Maar10} R. Maartens, Living Rev. Relativity \textbf{13}, 5 (2010).

\bibitem{Vile81} A. Vilenkin, Phys. Rev. D \textbf{23}, 852 (1981).

\bibitem{Gott85} J. R. Gott III, Astrophys. J. \textbf{288}, 422 (1985); W.
Hiscock, Phys. Rev. D \textbf{31}, 3288 (1985); B. Linet, Gen. Relativ.
Gravit. \textbf{17}, 1109 (1985); D. Garfinkle, Phys. Rev. D \textbf{32},
1323 (1985).

\bibitem{Kris97} A. Krishnan, et al, Nature \textbf{388}, 451 (1997); S.N.
Naess, A. Elgsaeter, G. Helgesen, and K.D. Knudsen, Sci. Technol. Adv.
Mater. \textbf{10}, 065002 (2009).

\bibitem{Grad} I. S. Gradshteyn and I. M. Ryzhik. \textit{Table of
Integrals, Series and Products} (Academic Press, New York, 1980)

\bibitem{Aram} E. R. Bezerra de Mello and A. A. Saharian, arXiv:1107.2557,
to appear in Class. Quantum Grav.

\bibitem{Saha11CP} A. A. Saharian and A. S. Kotanjyan, Eur. Phys. J. C
\textbf{71}, 1765 (2011).

\bibitem{Prud} A. P. Prudnikov, Yu. A. Brychkov and O. I. Marichev, \textit{%
Integrals and Series} (Gordon and Breach, New York, 1986), Vol. 2.

\bibitem{Spin} J. Spinelly and E.R. Bezerra de Mello, JHEP \textbf{12}
(2008) 081.

\bibitem{Abra72} \textit{Handbook of Mathematical Functions}, edited by M.
Abramowitz and I.A. Stegun (Dover, New York, 1972).

\bibitem{Aram2} E. R. Bezerra de Mello and A. A. Saharian, Class. and
Quantum Grav. \textbf{28}, 145008 (2011).

\bibitem{Birr82} N. D. Birrell and P. C. W. Davies, \textit{Quantum Fields
in Curved Space} (Cambridge University Press, Cambridge, 1982).

\bibitem{Burg85} C. P. Burgess and C. A. L\"{u}tken, Phys. Lett. B \textbf{%
153}, 137 (1985).

\bibitem{Camp91} R. Camporesi, Phys. Rev. D \textbf{43}, 3958 (1991).

\bibitem{Camp92} R. Camporesi and A. Higuchi, Phys. Rev. D \textbf{45}, 3591
(1992).

\bibitem{Kame99} M. Kamela and C. P. Burgess, Can. J. Phys. \textbf{77}, 85
(1999).

\bibitem{Cald99} M. M. Caldarelli, Nucl. Phys. B \textbf{549}, 499 (1999).

\bibitem{Knap04} A. Knapman and D. J. Toms, Phys. Rev. D \textbf{69}, 044023
(2004).

\bibitem{Saha05} A. A. Saharian, Nucl. Phys. B \textbf{712}, 196 (2005).

\bibitem{Saha06} A. A. Saharian, Phys. Rev. D \textbf{73}, 044012 (2006).

\bibitem{Saha06b} A. A. Saharian, Phys. Rev. D \textbf{73}, 064019 (2006).

\bibitem{SahaBook} A.A. Saharian, \textit{The Generalized Abel-Plana Formula
with Applications to Bessel Functions and Casimir Effect} (Yerevan State
University Publishing House, Yerevan, 2008); Preprint ICTP/2007/082;
arXiv:0708.1187.
\end{thebibliography}
\end{document}